\documentclass[copyright]{eptcs}
\providecommand{\event}{ACL2 2013}

\usepackage{breakurl}
\usepackage{color}
\usepackage{amsmath}
\usepackage{graphicx}
\usepackage{amsthm}
\usepackage{amssymb}
\usepackage{pstricks}
\usepackage{bbm}

\newcommand{\nocompile}[1]{}
\newcommand{\req}{\ensuremath{\mathtt{R}}}
\newcommand{\ack}{\ensuremath{\mathtt{A}}}
\newcommand{\select}{\mbox{select}}
\newcommand{\trans}[3]{#1\xrightarrow{#2}#3}

\newcommand{\inputwire}{\ensuremath{\mathtt{I}}}
\newcommand{\outputwire}{\ensuremath{\mathtt{O}}}

\DeclareMathOperator{\blocking}{blocking}

\DeclareMathOperator{\deadc}{\mathbf{Dead}}
\DeclareMathOperator{\idlec}{\mathbf{Idle}}
\DeclareMathOperator{\blockc}{\mathbf{Block}}

\newtheorem*{condition}{Complete Condition}
\newtheorem{runningexample}{Running Example, Part}

\theoremstyle{definition}
\newtheorem{lemma}{Lemma}
\newtheorem{theorem}{Theorem}

\begin{document}

\title{Verification of Building Blocks for Asynchronous Circuits}
\def\titlerunning{Verification of Building Blocks for Asynchronous Circuits}
\author{Freek Verbeek and Julien Schmaltz
\institute{Open University of The Netherlands \\ Heerlen, The Netherlands}
\institute{School of Computer Science}
\email{\{freek.verbeek,julien.schmaltz\}@ou.nl}
}
\def\authorrunning{Freek Verbeek and Julien Schmaltz}

\maketitle
\begin{abstract}
Scalable formal verification constitutes an important challenge for the design of asynchronous circuits.
Deadlock freedom is a property that is desired but hard to verify.
It is an emergent property that has to be verified monolithically.
We present our approach to using ACL2 to verify necessary and sufficient conditions over asynchronous delay-insensitive primitives.
These conditions are used to derive SAT/SMT instances from circuits built out of these primitives. These SAT/SMT instances help in establishing absence of deadlocks.
Our verification effort consists of building an executable checker in the ACL2 logic tailored for our purpose.
We prove that this checker is correct.
This approach enables us to prove ACL2 theorems involving {\tt defun-sk} constructs and free variables fully automatically.
\end{abstract}

\section{Introduction}

Today's hardware designs commonly are clocked.
A rhythmic clock signal ensures that a designer can assume a discrete notion of time.
The clocked design paradigm has many advantages, but they come at a high cost.
It induces overhead and delay in terms of speed, data flow and energy~\cite{sutherland12}.
In a clock-free or \emph{asynchronous} design each element acts only when necessary and at its own pace.
This can save energy, can increase speed and can decrease latency of communications.

Recently, Click has been proposed as a library for the design of asynchronous circuits~\cite{peeters10}.
It consists of primitives that are \emph{delay-insensitive}, i.e., primitives that behave correctly regardless of any delay induced by interfacing with the environment.
Click primitives are low-level hardware design templates for delay-insensitive elements such as storages, forks, joins and distributors~\footnote{To be more precise, the primitives are quasi-delay-insensitive. For sake of presentation, we do not distinct these terms.}.
Connected in a pipelined fashion, the purpose of these primitives is to behave as ``lego-like'' as possible.
They restore a high level of abstraction during the design phase, even when a close link to realistic asynchronous hardware is maintained.

Many state-of-the-art formal verification efforts on asynchronous circuits focus on proving properties over elements in isolation~\cite{zaki08,ouchet10,yan11}.
Deadlock freedom, however, is an emergent property.
Establishing deadlock freedom of primitives in isolation does not provide any information on deadlock freedom of the entire system.
A monolithic approach is mandatory. 
Our approach is to automatically derive SAT/SMT instances from Click circuits.
If the instance is infeasible, the circuit is deadlock-free.
If a solution is found, this solution corresponds to a structural deadlock.
This approach has been applied before to synchronous circuits, where it shows great promise in terms of scalability~\cite{verbeekschmaltz:fmcad10,gotmanov11}.

\begin{figure*}[!htb]
\centering
\scalebox{1}{
        \input{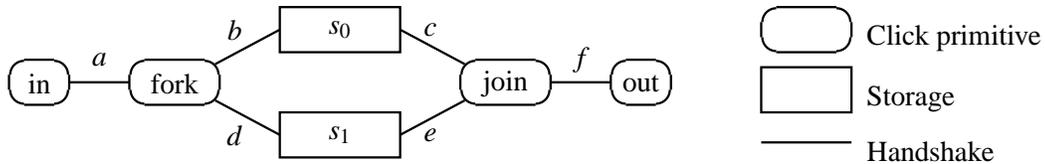}
}
\caption{Click circuit}
\label{fig:asynch_example}
\end{figure*}

Consider the network in Figure~\ref{fig:asynch_example} as an example.
The circuit is composed of six Click primitives.
These primitives use handshakes $a$ through $f$ to establish mutual communication.
The input injects packets which are duplicated by the fork. Two storages $s_0$ and $s_1$ buffer these packets.
The join waits for two packets at its inputs and combines them into one packet, which is sent to the output.

Given this circuit, we automatically derive the following result:
\begin{eqnarray}
\deadc(a)&\iff& ((s_0 \wedge \neg s_1) \vee (s_1 \wedge \neg s_0)) \wedge (s_0 = s_1)\label{eq:dead}
\end{eqnarray}
In words, this formula states that there is a deadlock involving handshake $a$ if and only if exactly one of the storages is full \emph{and} the internal state of both storages is equal.
The left hand side of the conjunct indicates that if, e.g., storage $s_0$ contains a packet but storage $s_1$ does not, a deadlock would occur.
In this configuration, the fork will never be able to duplicate two packets, whereas the join will never be able to combine two packets.
The right hand side indicates that \emph{invariably} both storages will either both be empty or both be full.
The formula is not satisfiable, i.e., there is no assignment of values to variables that makes the formula true.
Consequently, there is no deadlock.

Key to deriving a deadlock formula such as Equation~\ref{eq:dead}, is to establish necessary and sufficient conditions for each Click primitive in isolation.
These conditions must characterize the reasons that cause a handshake to be \emph{blocked}, i.e., not able to transmit a packet, or \emph{idle}, i.e., not receiving a packet.
We use the join as a running example.
The join waits for data from its two inputs $a$ and $b$ before forwarding data to output $c$.
Input $a$ is permanently blocked if and only if one of two cases arise.
First, when output~$c$ is permanently blocked the join can never forward a packet. It therefore blocks input $a$.
Secondly, when no packet arrives at input $b$, the join will never be able to merge two packets and will never produce an output. Input $a$ is blocked.
Hence the join induces the following necessary and sufficient condition:\\
\begin{condition}
The input of a join is permanently blocked if and only if either its output is permanently blocked or the other input is permanently idle.
\begin{eqnarray*}
\blockc(a) & \iff & \blockc(c) \vee \idlec(b)
\end{eqnarray*}
\end{condition}

Correct necessary and sufficient conditions for each primitive in the Click library are vital to the correctness of our approach.
Even though their correctness is often seems obvious, their formalizations are complicated and their proofs of correctness are often highly tedious.
Moreover, the Click library contains many primitives and our approach requires multiple necessary and sufficient conditions per primitive.
Therefore, we have implemented a small and highly tailored checker for Click primitives in ACL2.
This checker is able to automatically verify necessary and sufficient conditions built out of block- and idle predicates for a library of delay insensitive primitives.
This paper presents ACL2 details of our verification effort, a broader overview can be found in our publication at ASYNC~\cite{verbeekschmaltz:async13}.
Details on how these conditions can be used to build a formula such as Equation~\ref{eq:dead} can be found elsewhere~\cite{verbeekschmaltz:fmcad10,gotmanov11}.

\section{Formalizing Blocking and Idle Conditions}

We represent Click primitives using the eXtended Delay Insensitive (XDI) specification~\cite{verhoeff98,mallon98}.
In this paper, XDI specifications are represented using automata.
We first introduce the parts of the XDI formalism relevant to this paper.
Then, the execution semantics of XDI state machines are formalized.
Finally, we use Linear Temporal Logic (LTL) to formalize properties over executions of Click circuits.
LTL uses the {\tt G}(lobally) operator to express that some property is always true, and the {\tt F}(inally) operator which expresses that some property is eventually true~\cite{pnuelli77}.

\subsection{Formalization of Click Primitives}

A Click primitive is connected to several other primitives (its \emph{environment}) and may use several handshakes for this.
Each handshake is implemented by two wires $h_\req$ and $h_\ack$ for requests and acknowledgments.
Each wire is either an input to the primitive, or an output.
We allow the possibility that a request for handshake $h$ is accompanied by data $d$.
In this case, the handshake will be denoted with $h^d$.

\begin{runningexample}
For the join, the set of handshakes is $\{a,b,c\}$.
The set of input wires is $\{a_\req, b_\req, c_\ack\}$ and the set of output wires is $\{a_\ack, b_\ack, c_\req\}$.
The possibility of transmitting data with requests is not needed.
\end{runningexample}

\begin{figure}[!htbp]
\centering
\scalebox{1}{
        \input{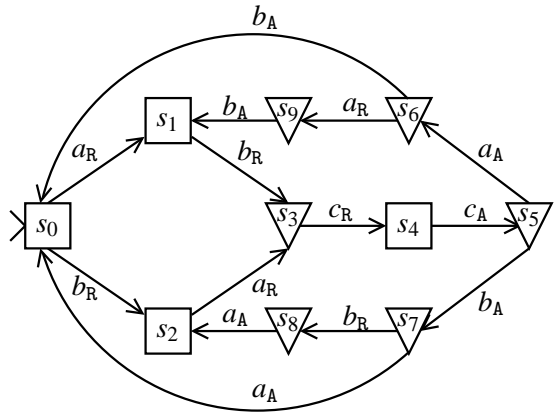}
}
\caption{XDI state graph of a join}
\label{fig:join_xdi}
\end{figure}
Figure~\ref{fig:join_xdi} shows the XDI state machine of the join.
The following ACL2 code, which will be explained in more detail hereafter, corresponds to this XDI state graph:
\begin{verbatim}
(defconst *xdi-sm-join*
 '(;;State Init Type      Transitions
     (s0   T    BOX       (((b R I) s2) ((a R I) s1))))
     (s1   NIL  BOX       (((b R I) s3)))
     (s2   NIL  BOX       (((a R I) s3)))
     (s3   NIL  TRANSIENT (((c R O) s4)))
     (s4   NIL  BOX       (((c A I) s5)))
     (s5   NIL  TRANSIENT (((b A O) s7) ((a A O) s6)))
     (s6   NIL  TRANSIENT (((b A O) s0) ((a R I) s9)))
     (s7   NIL  TRANSIENT (((a A O) s0) ((b R I) s8)))
     (s8   NIL  TRANSIENT (((a A O) s2)))
     (s9   NIL  TRANSIENT (((b A O) s1))))
\end{verbatim}

An XDI specification consists of a set of states.
There is exactly one state that is the initial state.
In contrast to the full XDI specification, which provides five different types of states, our presentation allows only two types of states: \emph{indifferent} states (denoted with $\mathtt{BOX}$) and \emph{transient} states.
An indifferent state poses no progress obligation on either the circuit or its environment.
A transient state requires progress of the circuit, i.e., the primitive eventually has to proceed to a next state.

\begin{runningexample}
For the join, state $s_0$ is the initial state.
As this state requires an input from the environment on wires $a_\req$ and $b_\req$, there are no progress obligations and the type of state $s_0$ is indifferent.
In state $s_3$, requests have been received from both $a$ and $b$.
The circuit has to send a request to its output $c$. Consequently, the type of state $s_3$ is transient.
\end{runningexample}

A transition {\tt (w s)} is a tuple containing the label {\tt w} that represents the wire on which a communication is to occur for the transition to the next state {\tt s} to happen.
A wire {\tt w} is represented by a tuple {\tt (h R/A I/O)} with three values representing the handshake, whether the wire is used for Requests or Acknowledgments, and whether the wire is an input or and output to the primitive.
For example, if the join is in its initial state and the input wire $a_\req$ changes from low to high, the join moves to state $s_1$.
For details on rules on which transitions are allowed and required in XDI specifications, we refer to papers on the XDI formalism (e.g.,~\cite{verhoeff98}).

\subsection{Execution Semantics}

The execution semantics of an XDI state machine $X$ are formalized relative to its environment.
Since the environment consists of Click primitives, it is basically a large XDI state machine.
The only information relevant to the analysis of primitive $X$, is whether its input wires are \emph{stable} or not.
A wire $w$ is stable if and only if its value is permanently unchanged. This implies that if wire $w$ is stable, no transition labelled with~$w$ occurs.
Therefore, the environment, i.e., the complete set of Click primitives constituting the circuit, is represented as a set of input wires such that each wire in the set is deemed to be stable.

\begin{runningexample}\label{rex:env}
We consider an environment of the join in which wire $c_\ack$ is stable.
Any execution will strand in state $s_4$, i.e., waiting for the environment to acknowledge the receipt of data by output $c$ after a request to $c$ has been sent to fetch this data.
Essentially, the join is dead because the environment permanently \emph{blocks} on handshake $c$.
In another environment wires $a_\req$ and $b_\req$ may be stable.
The join will get stuck in its initial state $s_0$.
It is waiting for the environment to send requests.
Essentially, the join is dead because the environment is permanently \emph{idle} on handshakes $a$ and $b$.
In total, the three input wires induce $2^3$ different possible groups of environments of interest while analyzing the join, ranging from a live one (i.e., the environment is the empty set), to an environment where all three input wires of
the join are stable (i.e., the environment is the set $\{a_\req, b_\req, c_\ack\}$).
\end{runningexample}

First, we define a predicate to indicate that a given wire is an input wire.
We make use of the {\tt proj} function, which returns the $n$th projection in a list of lists.
For example, {\tt (proj 0 *xdi-sm-join*)} return the set of states of the XDI state machine of the join.
\begin{verbatim}
(defun input-wirep (xdi-sm wire)
  (member-equal (list (car wire) (cadr wire) 'I)
                (proj 0 (union (proj 3 xdi-sm)))))  
\end{verbatim}
Function {\tt input-wirep} takes as input an XDI state machine and a partial description of a wire, namely a tuple {\tt(h R/A)}, where {\tt h} is the handshake and {\tt R/A} indicates a Request or Acknowledge.
This is transformed to a wire {\tt(h R/A I)}, for which is searched in the set of labels on the transitions of the XDI state machine.
An environment can now be defined as a set of input wires.
\begin{verbatim}
(defun envp (xdi-sm env)
  (if (endp env)
    t
    (and (input-wirep xdi-sm (car env))
         (envp xdi-sm (cdr env)))))
\end{verbatim}
The actual value of the environment depends on the state of the network.
In the remainder of this paper, we quantify over \emph{all} possible environments.
\\

Given an environment, we can define the next step function of XDI state machines.
Given a current state, this function returns the set of next possible states.
First, we define a function that takes as input a list of transitions {\tt ts} and filters out transitions labelled with a stable wire, i.e., transitions that cannot occur.
Function {\tt stable} returns {\tt t} if and only if the given wire is stable in the given environment.
\begin{verbatim}
(defun remove-stable-wire-transitions (ts env)
  (cond ((endp ts)
         nil)
        ((stable (caar ts) env)
         (remove-stable-wire-transitions (cdr ts) env))
        (t
         (cons (car ts)
               (remove-stable-wire-transitions (cdr ts) env)))))
\end{verbatim}
This function yields all transitions {\tt((h R/A I/O) s')} that are not stable.
The next step function gives it all possible transitions from the current state {\tt s} and takes from each resulting transition the next state~{\tt s'}.
\begin{verbatim}
(defun xdi-step (xdi-sm s env)
  (proj 1 (remove-stable-wire-transitions (nth 3 (assoc s xdi-sm)) env)))
\end{verbatim}

\subsection{Labelling States as Blocking or Idling}

We identify each non-transient state as \emph{blocking} or \emph{idling} with respect to handshake $h$.
We define these labels in such a way that if primitive $X$ is permanently stuck in a state labelled as ``blocking $h$'', handshake $h$ is permanently blocked.
Handshake $h$ is permanently idle, if primitive $X$ permanently remains in a state labelled ``idling $h$''.

\begin{runningexample}
For the join, we consider handshake $a$, which uses wires $a_\req$ and $a_\ack$ to communicate with the join.
States $s_1$, $s_3$, $s_4$, $s_5$, $s_7$, $s_8$ and $s_9$ are blocking this handshake.
In these states, handshake $a$ has sent a request to the join, which has not been acknowledged by the join yet.
If the join is permanently stuck in these states, handshake $a$ will permanently wait for an acknowledgment from the join.
Handshake $a$ is permanently blocked.
The remaining states $s_0$, $s_2$, and $s_6$ are idling handshake $a$.
In these states, the join waits for a request from handshake $a$.
When it is permanently stuck in of these states, handshake $a$ is failing to send this request.
Handshake $a$ is permanently idle.
\end{runningexample}

To define predicates {\tt blocking} and {\tt idling}, we define an executable function {\tt compute-b/i} which recursively explores the state machine and returns an association list mapping to each state a Boolean value indicating how the state should be labelled.
The intuition of this function is that initially states are idling.
As soon as a transition {\tt((h R I/O) s')} occurs, apparently the primitive is in a state where it has been requested to communicate on handshake {\tt h}, but has not finished this communication yet.
All subsequent states are therefore blocking handshake $h$, until a transition {\tt((h A I/O) s')} occurs.
After this transition, the primitive has successfully dealt with the request and no communication occurs on handshake $h$.
All subsequent states are idling handshake $h$.
This repeats, until all states have been explored.

\begin{figure}
\begin{verbatim}
(mutual-recursion
 (defun compute-b/i (xdi-sm s h flg ret) 
   (if (assoc s ret)
     ret
     (let ((ret (acons s flg ret)))
       (compute-b/i-ts xdi-sm (nth 3 (assoc s xdi-sm)) h flg ret))))
 (defun compute-b/i-ts (xdi-sm ts h flg ret)
   (let ((ret (cond ((equal (caaar ts) h)
                     (compute-b/i xdi-sm (cadar ts) h (not flg) ret))
                    (t
                     (compute-b/i xdi-sm (cadar ts) h flg ret)))))
     (compute-b/i-ts xdi-sm (cdr ts) h flg ret))))
\end{verbatim}
\caption{Implementation of {\tt compute-b/i}}
\label{fig:compute-bi}
\end{figure}
Figure~\ref{fig:compute-bi} shows the ACL2 code of function {\tt compute-b/i} which computes this association list.
Function {\tt compute-b/i} takes as second parameter a state {\tt s}.
The third parameter is a flag indicating whether currently explored states are to be marked blocking or idling.
It checks whether this state has already been explored. 
If so, then no further exploration is needed.
Otherwise, it updates the returned association list {\tt ret} by associating the current value of the flag to the current state.
After this update, the function recursively explores all transitions leading out of the current state.
Function {\tt compute-b/i-ts} takes as second parameter a set of transitions.
Sequentially two things occur.
First, the first transition of the set is analyzed.
If this transition concerns handshake {\tt h}, the flag is changed indicating that a switch from blocking to idling happens, or the other way around.
A recursive call with the next state as value for {\tt s} is performed.
Second, the remaining transitions are recursively explored.

Function {\tt compute-b/i} is initially called with the initial state and as flag the value {\tt nil}. 
The predicate {\tt blocking} can now be defined by simply looking up the given state in the result of function {\tt compute-b/i}.
\begin{verbatim}
(defun blocking (xdi-sm s h)
  (cadr (assoc s (compute-b/i xdi-sm (xdi-get-init-state xdi-sm) h nil nil))))
\end{verbatim}
Predicate {\tt idling} is defined as not blocking.

\begin{runningexample}
The state machine of the join contains a transition $\trans{s_0}{a_\req}{s_1}$.
We can compute that:\\
{\tt (blocking *xdi-sm-join* s0 a)}\\
evaluates to {\tt nil}, whereas\\
{\tt (blocking *xdi-sm-join* s1 a)}\\
evaluates to~{\tt t}.\\
This represents that state $s_0$ is idling handshake $a$, whereas state $s_1$ is blocking handshake $a$.
\end{runningexample}

\subsubsection*{Remarks}

An important assumption on the Click primitives is that their XDI specification ensures that function {\tt blocking} is uniquely defined over all non-transient states.
If a certain state $s$ can be reached from the initial state using a sequence of transitions with one transition labelled $h_\req$ and no transition labelled $h_\ack$, function {\tt blocking} enforces $\blocking(h, s)$ to be true.
If this state $s$ can also be reached with a sequence of transitions without $h_\req$ as label, function {\tt blocking} enforces $\blocking(h, s)$ to be false.
Such state graphs are not allowed.
We will call a Click primitive for which function $\blocking$ is unique over all non-transient states \emph{unambiguous}.
In ACL2, we have an executable function checking for unambiguity.

Function {\tt compute-b/i} does not necessarily terminate.
To prove termination, we require both a list of assumptions and some checks which have to be performed by the function before each recursive call.
We have added the assumption as guards, and defined a logical version of this function with the additional checks.
For the logical version, we have proven termination.
The code shown here is the executable version, without these checks.
Using an {\tt mbe}-construct, we have proven that under assumption of the guards, the logical and the executable versions are  equivalent.

\subsection{Formulating Block- or Idle Conditions}

Whether a primitive can be stuck in a blocking- or idling state depends on the environment.
Consider again the state machine of the join (see Figure~\ref{fig:join_xdi}).
If the environment dictates that wire $c_\ack$ is stable, any execution will strand in state $s_4$.
This state is blocking handshake $a$ and therefore handshake $a$ is permanently blocked.
We say that a handshake $h$ is permanently blocked if and only if a primitive will eventually get stuck in non-transient states labelled ``blocking $h$''.

To define LTL properties over XDI state machines, we first define the notion of trace.
A trace is a set of states that is connected via the {\tt xdi-step} function.
\begin{verbatim}
(defun xdi-tracep (xdi-sm trace env)
  (cond ((endp trace)
         t)
        ((endp (cdr trace))
         t)
        (t
         (and (member (cadr trace) (xdi-step xdi-sm (car trace) env))
              (xdi-tracep xdi-sm (cdr trace) env)))))
\end{verbatim}
To express that a machine is permanently stuck in blocking states, we use a {\tt defun-sk} construct to quantify over all possible traces starting in the current state.
\begin{verbatim}
(defun-sk G-blocking_ (xdi-sm h s env)
  (forall (trace)
          (implies (and (xdi-tracep xdi-sm trace env)
                        (equal (car trace) s))
                   (or (equal (nth 2 (assoc (car (last trace)) xdi-sm))
                              'transient)
                       (blocking xdi-sm (car (last trace)) h)))))
\end{verbatim}
The trailing underscore is used to indicate that the function is non-executable.
Any trace starting in {\tt s} ends either in a transient state or in a state that is blocking handshake {\tt h}.
Note that we deal with finite traces only.
Since the XDI automata are always finite, any infinite trace consists of a prefix followed by a repetition of some trace induced by a cycle.
It it therefore sufficient to analyze all finite -- but of unbounded length -- prefixes.

Similarly, we express the {\tt F} operator using a {\tt defun-sk} construct introducing an existential quantifier.
\begin{verbatim}
(defun-sk F-G-blocking_ (xdi-sm h s env)
  (exists (trace)
          (and (xdi-tracep xdi-sm trace env)
               (equal (car trace) s)
               (G-blocking_ xdi-sm h (car (last trace)) env))))
\end{verbatim}
Similar definitions have been formulated for idling.
Given environment {\tt env},  handshake $h$ is \emph{permanently blocked} if and only if the corresponding XDI state machine is eventually always in a blocking state.
Similarly, handshake $h$ is \emph{permanently idle} if and only if the corresponding XDI state machine is eventually always in an idling state.
\begin{verbatim}
(defun Blocked_ (xdi-sm h env)
  (F-G-blocking_ X h (xdi-get-init-state xdi-sm) env))
(defun Idle_ (xdi-sm h env)
  (F-G-idling_ xdi-sm h (xdi-get-init-state xdi-sm) env))
\end{verbatim}

Finally, we can formulate necessary and sufficient conditions per Click primitive.
For example, the ACL2 formalization of the running example becomes:
\begin{theorem}\label{thm:block_join}\hspace{1ex}
\begin{verbatim}
(defthm blocking-equation-join
  (implies (envp xdi-sm env)
           (iff (Blocked_ *xdi-sm-join* 'a env)
                (or (Blocked_ *xdi-sm-join* 'c env)
                    (Idle_ *xdi-sm-join* 'b env)))))
\end{verbatim}
\end{theorem}

\section{Model Checking Blocking and Idle Conditions}

Proving Theorem~\ref{thm:block_join} could be done manually as follows. First, a case distinction is required over all possible environments (in this case, eight in total).
For each environment, all traces have to be explored to see whether the labels computed by function {\tt compute-b/i} always satisfy the formula that is to be proven.
As we need multiple theorems per primitive and there is a whole library of Click primitives, we want to prove dozens of theorems such as Theorem~\ref{thm:block_join}.
Naturally, a manual proof is simply infeasible and an automated approach is mandatory.
Therefore, our proof technique is to A.) define executable counterparts to non-executable functions {\tt Blocked\_} and {\tt Idle\_} and B.) make an automatic enumeration of all possible environments.
A once and for all proof that these functions are correctly implemented can then be used to prove Theorem~\ref{thm:block_join} without further interaction.

\subsection{ACL2 Overview of Automated Proof}

Our objective for Part A. is to implement function {\tt Blocked} in such a way that the following lemma can be proven (similar for {\tt Idle}). This is the specification of function {\tt blocked}, its definition will follow.
\begin{lemma}\label{lem:rewrite_block}\hspace{1ex}
\begin{verbatim}
(defthm rewrite-non-exec-Blocked_-to-exec-Blocked
  (implies (and (xdi-smp xdi-sm)
                (member s (proj 0 xdi-sm)))
           (equal (Blocked_ xdi-sm h s env)
                  (Blocked  xdi-sm h s env))))
  :rule-classes :definition)
\end{verbatim}
\end{lemma}
Function {\tt xdi-smp} recognizes syntactically valid XDI state machines.
For any state {\tt s} that is a valid state, the result of function {\tt Blocked} is equivalent to that of its specification {\tt Blocked\_}.
Once this theory has been established, the non-executable definition {\tt Blocked\_} is disabled in the theory.
\begin{verbatim}
(in-theory (disable Blocked_))
\end{verbatim}
This way we are sure -- when proving a theorem -- that any occurrence of {\tt Blocked\_} will be rewritten using Lemma~\ref{lem:rewrite_block} \emph{only} (note that a defun-sk construct is non-executable, but is still often rewritten to a goal without the original function name, thereby preventing application of Lemma~\ref{lem:rewrite_block}).

As for Part B., we straightforwardly implement a function {\tt reasonable-envs} which takes as input an XDI state machine and generates a list of all possible environments.
We first implement function {\tt compute-input-wires} which given an XDI state machine returns the set of input wires.
The set of environments is then computed as follows:
\begin{verbatim}
(defun reasonable-envs (xdi-sm)
  (powerset (remove-duplicates (compute-input-wires xdi-sm))))
\end{verbatim}
The set of input wires is assembled. Duplicate entries are removed, for sake of efficiency. The list of relevant environments contains any subset of these wires.

\subsubsection*{Remark}
The number of environments grows exponentially.
The XDI automata of interest are however not very large.
In Section~\ref{sec:app} we apply our method to a non-trivial Click primitive, namely the distributor.
Regardless of the large number of environments, we can easily deal with this primitive.
\\

Using Parts A. and B., we can prove Theorem~\ref{thm:block_join} completely automatically, after rewriting it slightly.
Our final formulation becomes:
\begin{verbatim}
(defthm blocking-equation-join
  (and (xdi-smp-guard *xdi-sm-join*)
       (implies (member env (reasonable-envs *xdi-sm-join*))
                (iff (Blocked_ *xdi-sm-join* nil 'in0 env)
                     (or (Blocked_ *xdi-sm-join* nil 'out env)
                         (Idle_ *xdi-sm-join* nil 'in1 env))))))
\end{verbatim}

First, we explicitly verify that the constant {\tt *xdi-sm-join*} satisfies all guards necessary for correct execution of functions {\tt Blocked} and {\tt Idle}. 
Function {\tt xdi-smp-guard} is executable and therefore {\tt (xdi-smp-guard *xdi-sm-join*)} is proven without further interaction.
Secondly, we reformulate the theorem, so that open variable {\tt env} is a member of a computable set of environments.
The ACL2 simplifier will compute all reasonable environments.
Subsequently, having the following lemma enabled in the theory ensures that the {\tt member} construct breaks the goal down into eight different subgoals (one for each environment):
\begin{verbatim}
(defthm member-rewrite
  (equal (member a (cons b x))
         (if (equal a b) (cons b x)
             (member a b))))
\end{verbatim}
For each subgoal, the ACL2 simplifier uses Lemma~\ref{lem:rewrite_block} (and a similar lemma for {\tt Idle}) to rewrite the subgoal to executable versions of {\tt Blocked} and {\tt Idle}.
At this point, all functions are executable and there are no variables.
The truth of each subgoal is automatically evaluated.

\subsection{Implementation of {\tt Blocked} and {\tt Idle}}

The implementation of {\tt Blocked} needs to check whether eventually generally the machine is in states that are either transient or labelled as ``blocking'' by function {\tt compute-b/i}.
So first a function must be implemented which decides whether in a certain start state {\tt s} the machine generally is in such states, i.e., a function {\tt G-blocking} must be implemented.
Figure~\ref{fig:G-blocking} shows the implementation of this function.
\begin{figure}[ht!]
\begin{verbatim}
(mutual-recursion
 (defun G-blocking (xdi-sm visited s h env)
   (if (not (member-equal s visited))
     (if (or (equal (nth 2 (assoc s xdi-sm)) 'transient)
             (blocking xdi-sm s h))
       (G-blocking-ss xdi-sm (cons s visited)
                      (remove-equal s (xdi-step xdi-sm s env)) h env)
       nil)
     visited))
 (defun G-blocking-ss (xdi-sm visited ss h env)
   (if (endp ss)
     visited
     (let ((ret (G-blocking xdi-sm visited (car ss) h env)))
       (cond ((equal ret nil)
              ret)
             (t
              (let ((ret2 (G-blocking-ss xdi-sm visited (cdr ss) h env)))
                (if ret2
                  (append ret ret2)
                  nil))))))))

\end{verbatim}
\caption{Implementation of {\tt G-blocking}}
\label{fig:G-blocking}
\end{figure}

Function {\tt G-blocking} takes as input the XDI state machine, an accumulator of visited states, the current state, the handshake and the current environment.
If it returns {\tt nil} this indicates that there is some reachable state that is not labelled ``blocking'' with respect to handshake {\tt h}.
If it does not return {\tt nil} it does not return {\tt t}, but instead it returns a list of states that has been explored.
This will be used in the proofs later on.

If the current state {\tt s} is not visited and labelled ``blocking'' with respect to handshake {\tt h}, exploration continues with all next states using function {\tt G-blocking-ss}.
This function deals with two cases: either there are no states to be explored, or there are states to be explored.
In the first case, as there are no more reachable states, all reachable states have been explored.
Therefore, the function should return {\tt t}, but instead returns the accumulator {\tt visited} storing all explored states.
Otherwise, the function checks whether the first state in the list is {\tt G-blocking}. 
If the result of this check is {\tt nil}, this result is returned.
Otherwise, the intermediate result (i.e., the states explored in the recursive call) are appended to the final result.

A similar function is implemented to compute {\tt F-G-blocking\_}.
We extend both functions with an extra flag so that these functions can also be used to compute {\tt (F-)G-idling}.
Using these functions, we can define the implementations of {\tt Blocked\_} and {\tt Idle\_}.
\begin{verbatim}
(defun Blocked (xdi-sm h env)
  (consp (F-G-blocking xdi-sm nil `(,(xdi-get-initial-state xdi-sm)) h env)))
(defun Idle (xdi-sm h env)
  (consp (F-G-idling xdi-sm nil `(,(xdi-get-initial-state xdi-sm)) h env)))
\end{verbatim}

\subsection{Proof of Lemma~\ref{lem:rewrite_block}}

We present the proof of correctness of function {\tt G-blocking}. The proofs for {\tt F-G-blocking} are similar.
The proof is in two directions.
We prove Lemma~\ref{lem:spec-->exec} which states that for any state such that specification {\tt G-blocking\_} returns {\tt t}, executable function {\tt G-blocking} returns a non-empty list of visited states.
Secondly, we prove Lemma~\ref{lem:exec-->spec} which states that any state for which executable function {\tt G-blocking} returns a non-empty list, specification {\tt G-blocking\_} returns {\tt t}.\\

\begin{lemma}\label{lem:spec-->exec}\hspace{1ex}

We formulate the lemma in such a way that induction over {\tt G-blocking} is possible.
This requires parameters {\tt visited} and {\tt ss} to be free.
Also, any assumption on these variables must be an invariant over function {\tt G-blocking}.
Our formalization is as follows, and will be detailed hereafter:
\begin{verbatim}
(defthm G-blocking_-->G-blocking
  (implies (and (xdi-smp xdi-sm)
                (G-blocking_ xdi-sm h s env)
                (consp ss)
                (subsetp ss (proj 0 xdi-sm))
                (A-xdi-reachable xdi-sm env s ss))
           (G-blocking xdi-sm visited ss h t env)) 
\end{verbatim}
Assume a valid XDI state machine and a state {\tt s} which is generally blocking according to the specification.
For any non-empty set {\tt ss} of valid states we prove that executable function {\tt G-blocking} returns a non-empty list.
As an invariant, we require that all states in {\tt ss} are reachable from state {\tt s}.
Function {\tt A-xdi-reachable} returns {\tt t} if and only if all states in {\tt ss} are reachable from state {\tt s}.
Reachability between two states {\tt s1} and {\tt s2} is straightforwardly defined using a {\tt defun-sk} construct as the existence of a non-empty trace starting in {\tt s1} and ending in {\tt s2}:
\begin{verbatim}
(defun-sk xdi-reachable (xdi-sm env s1 s2)
  (exists (trace)
          (and (xdi-tracep xdi-sm trace env)
               (consp trace)
               (equal (car trace) s1)
               (equal (car (last trace)) s2))))
\end{verbatim}
Once it is proven that reachability of states {\tt ss} from state {\tt s} is indeed an invariant, the proof becomes conceptually very easy.
As soon as G-blocking encounters a non-transient state {\tt s2}, we know from the invariant that state {\tt s2} is reachable from state {\tt s}.
From assumption {\tt (G-blocking\_ xdi-sm h s env)} we can prove that {\tt s2} is blocking, which is expressed by lemma {\tt spec-of-G-blocking\_}: 
\begin{verbatim}
(defthm spec-of-G-blocking_
  (implies (and (G-blocking_ xdi-sm h s env)
                (xdi-reachable xdi-sm env s s2)
                (not (equal (nth 2 (assoc s2 xdi-sm)) 'transient)))
           (blocking xdi-sm s2 h))
  :hints (("Goal" :use ((:instance G-blocking_-necc
                         (trace (xdi-reachable-witness xdi-sm env s s2)))))))
\end{verbatim}
The theorem is proven by instantiating the theorem introduced by the {\tt defun-sk} event corresponding to {\tt G-blocking\_}.
This instantiation requires a trace from {\tt s} to {\tt s2}.
This trace is exactly the witness created by the {\tt defun-sk} event corresponding to {\tt xdi-reachable}.

Now we do induction over {\tt G-blocking}.
We have to prove that it does not return {\tt nil}.
It returns {\tt nil} only if it encounters an illegal state or a non-transient state that is not labelled ``blocking''.
The first case cannot happen due to assumption {\tt (xdi-smp xdi-sm)}.
The second case cannot happen since the invariant ensures that any explored state is reachable from state {\tt s} and since lemma {\tt spec-of-G-blocking\_} can be used to prove that any state reachable from {\tt s} is labelled ``blocking''.\\
\end{lemma}

\begin{lemma}\label{lem:exec-->spec}\hspace{1ex}

The lemma is formulated as follows:
\begin{verbatim}
(defthm G-blocking-->G-blocking_
  (implies (and (xdi-smp xdi-sm (cars xdi-sm))
                (member-equal s (cars xdi-sm))
                (G-blocking xdi-sm nil (list s) h env))
           (G-blocking_ xdi-sm h s env)))
\end{verbatim}
Given a valid XDI state machine and a valid state {\tt s}, the executable function correctly decides the LTL formula for state {\tt s}.

For the proof of the lemma we use the fact that function {\tt G-blocking} does not return {\tt t} when all reachable non-transient states are labelled ``blocking'', but the accumulator {\tt visited} instead. 
As we reason over reachable states, we define function {\tt xdi-reach} which takes as input a set of states {\tt ss} and assembles all states reachable from any state in {\tt ss}.
The reasoning is as follows:
\begin{enumerate}
\item Any state returned by {\tt G\_blocking} is either transient or labelled ``blocking''.
\item For any set of states {\tt ss}, the set of states accumulated by \\
{\tt (G-blocking xdi-sm visited ss h env)} contains all states returned by {\tt (xdi-reach ss)}.
\item Any state reachable from any state in {\tt ss} is in {\tt (xdi-reach ss)}.
\item We have to prove that any state {\tt s2} reachable from {\tt s} is either transient or labelled ``blocking''.\\
 By ~3.) it follows that:\\
 {\tt (member s2 (xdi-reach (list s))}\\
 By 2.) it follows that:\\
 {\tt (member s2 (G-blocking xdi-sm nil (list s) h env)}\\
 By 1.) it follows that {\tt s2}  is either transient or labelled ``blocking''.
\end{enumerate}

First, we prove a theorem stating that any non-transient state returned by {\tt G-blocking} is indeed labelled ``blocking''.
\begin{verbatim}
(defthm all-states-in-G-blocking-are-blocking
  (implies (A-blocking xdi-sm visited h)
           (A-blocking xdi-sm (G-blocking xdi-sm visited ss h env) h)))
\end{verbatim}
Function {\tt A-blocking} is a universal quantifier expressing that all given states are either transient or labelled ``blocking''.
Assuming this property holds for all initially accumulated states, this property holds for accumulated states.

We then prove that {\tt G-blocking} returns all states assembled by {\tt xdi-reach}.
\begin{verbatim}
(defthm all-reachable-states-in-G-blocking
  (implies (G-blocking xdi-sm visited ss h env) 
           (subsetp (xdi-reach xdi-sm visited ss env)
                    (G-blocking xdi-sm visited ss h env))))
\end{verbatim}
Assuming that {\tt G-blocking} does not return {\tt nil}, we prove that any state that is in the reach of some state in {\tt ss} is also member of the list of accumulated states returned by {\tt G-blocking}.

Finally, we prove correctness of function {\tt xdi-reach}, i.e., that it contain all reachable states.
\begin{verbatim}
(defthm spec-of-xdi-reach
  (implies (and (xdi-smp xdi-sm (proj 0 xdi-sm))
                (member s (proj 0 xdi-sm))
                (xdi-reachable xdi-sm env s s2))
           (member-equal s2 (xdi-reach xdi-sm nil (list s) env)))
\end{verbatim}

Using these lemmas, Lemma~\ref{lem:exec-->spec} can be proven without induction.
To prove the universal quantifier introduced by {\tt G-blocking\_}, we have to prove of a witness state
\begin{center}
$s_w$ = {\tt (G-blocking\_-witness xdi-sm h s env)}
\end{center}
that it is either transient or labelled ``blocking''.
Instantiating the first lemma with the accumulator {\tt visited} set to {\tt nil}, automatically discharges its assumption.
What remains to be proven is that $s_w$ is accumulated by {\tt G-blocking}.
This is proven by the second lemma, instantiated with {\tt visited} set to {\tt nil} and {\tt ss} set to {\tt (list s)}.
This forces us to prove that state $s_w$ is a member of {\tt xdi-reach}.
The third lemma is used to prove this, instantiating {\tt s2} with {\tt (G-blocking\_-witness xdi-sm h s env)}.
\end{lemma}

\section{Application}\label{sec:app}

\begin{figure}[h!]
\centering
\scalebox{1}{
        \input{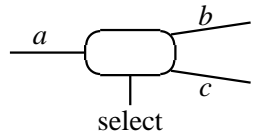}
}
\caption{Schematic overview of the distributor}
\label{fig:dist_overview}
\end{figure}

\begin{figure*}
\centering
\resizebox{\hsize}{!}{
	\input{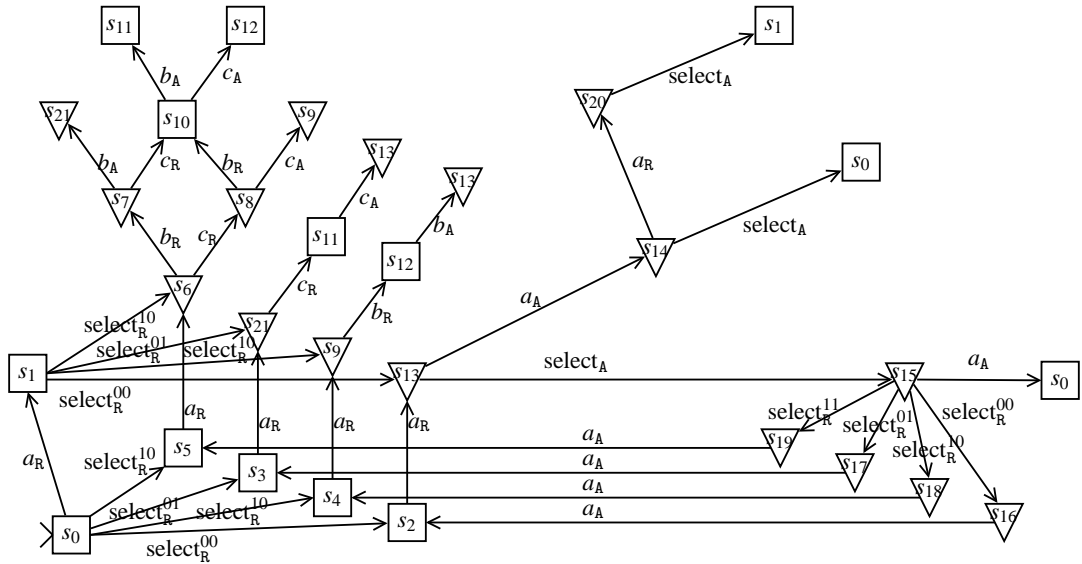}
}
\caption{XDI state graph of the distributor. Note that several states (e.g. $s_{11}$ and $s_{12}$ are replicated in order to simplify the diagram.}
\label{fig:dist_xdi}
\end{figure*}

The \emph{distributor} (see Figure~\ref{fig:dist_overview}) is a Click primitive used for routing packets through a network.
It uses handshake $a$ on which the availability of data is communicated and three handshakes $\select^d$ ($00 \leq d \leq 11$).
If $d = 00$, the incoming data is dropped.
If $d = 01$, the packet is routed towards output $b$.
Similarly $d = 10$ routes towards output $c$.
Figure~\ref{fig:dist_xdi} shows the XDI state graph of the distributor.
The set of input wires $W_\inputwire$ is  $\{a_\req, \select^d_\req, b_\ack, c_\ack\}$.
The remaining wires are output, i.e., $W_\outputwire = \{a_\ack, \select_\ack, b_\req, c_\req\}$.
Handshake $\select$ uses data for requests.

The blocking equation of input {\tt a} of the distributor is shown in Figure~\ref{fig:block_dist}.
Blockage of input {\tt a} is logically equivalent to three cases.
First, if no select signal ever arrives at input {\tt select}, the distributor will not know how to route packets and will therefore not transmit them.
Secondly, if always eventually a $01$ signal arrives on the {\tt select} wire, 	and if output {\tt b} is permanently blocked, eventually a packet at input {\tt a} will be permanently blocked (note that a $01$ signal on the {\tt select} wire means ``route towards output {\tt b}'').
The third case is similar but for output {\tt c}.

Using the lemmas presented in the previous section, theorem in Figure~\ref{fig:block_dist} can be proven instantaneously and without any interaction.
\begin{figure}
\begin{verbatim}
(defthm blocking-equation-distributor
  (and (xdi-smp-guard *xdi-sm-distributor*)
       (implies (member-equal env (reasonable-envs *xdi-sm-distributor*))
                (iff (Blocked_ *xdi-sm-distributor* 'in env)
                     (or (and (Idle_ *xdi-sm-distributor* 'select00 env)
                              (Idle_ *xdi-sm-distributor* 'select01 env)
                              (Idle_ *xdi-sm-distributor* 'select10 env))
                         (and (not (Idle_ *xdi-sm-distributor* 'select01 env))
                              (Blocked_ *xdi-sm-distributor* 'out1 env))
                         (and (not (Idle_ *xdi-sm-distributor* 'select10 env))
                              (Blocked_ *xdi-sm-distributor* 'out0 env)))))))
\end{verbatim}
\caption{Blocking Equation for Distributor}
\label{fig:block_dist}
\end{figure}

\section{Conclusion}

We have mechanically verified properties of a library of delay-insensitive primitives in the ACL2 theorem prover.
These properties are often deceptively simple, making it easy to formulate incorrect theorems.
Moreover, their proofs are large and cumbersome.
Their formalization is tricky: it is based on XDI state machines and their execution semantics relative to the environment of the primitive.
Our approach consists of building a checker for XDI state machines which can decide LTL formulae that are built out of block- and idle predicates.
This checker has been proven correct with respect to its specification.
The theorems that are to be proven involve free variables and non-executable functions introduced by {\tt defun-sk} constructs.
Loading the book that contains our definitions and lemmas suffices to fully automatically prove these theorems quickly.

The properties that have been proven are used to derive a SAT/SMT instance from an asynchronous circuit built out of primitives in the library.
This derivation can be quite contrived, especially when data is taken into account.
In the future, we plan to use ACL2 to prove correctness of our derivation, proving that feasibility of the derived SAT/SMT instance is logically equivalent to the existence of a structural deadlock.

\bibliographystyle{eptcs}
\bibliography{ref} 

\end{document}